
\input phyzzx
\hsize=6.5in
\hoffset=0.0in
\voffset=0.0in
\vsize=8.9in
\FRONTPAGE
\line{\hfill BROWN-HET-914}
\line{\hfill July 1993}
\vskip1.5truein
\titlestyle{{CLASSICAL AND QUANTUM THEORY OF PERTURBATIONS IN INFLATIONARY
UNIVERSE MODELS}\foot{Invited
lecture presented by R.B. at the 37th Yamada Conference 'Evolution of the
Universe and its Observational Quest', June 7 - 12 1993, to be publ. in the
proceedings, ed. by K. Sato (Universal Academy Press, Tokyo, 1993).}\break}
\bigskip
\author{R. Brandenberger$^{1)}$, H. Feldman$^{2)}$ and V.
Mukhanov$^{3)}$\foot{On leave of absence from Institute for Nuclear Research,
Academy of Sciences, Moscow 117 312, Russia}}
\centerline{1) {\it Department of Physics}}
\centerline{{\it Brown University, Providence, RI 02912, USA}}
\medskip
\centerline{2){\it Physics Department}}
\centerline{\it University of Michigan, Ann Arbor, MI 48109}
\medskip
\centerline{3){\it Institute for Theoretical Physics}}
\centerline{\it ETH H\"onggerberg, CH-8093 Z\"urich, Switzerland}
\bigskip
\abstract
A brief introduction to the gauge invariant classical and quantum
theory of cosmological perturbations is given.  The formalism is
applied to inflationary Universe models and yields a consistent and
unified description of the generation and evolution of fluctuations.
A general formula for the amplitude of cosmological perturbations in
inflationary cosmology is derived.
\endpage
\noindent \undertext{1.  Introduction}:
\par
According to the cosmological principle, the Universe should be
homogeneous on large scales.  The isotropy of the cosmic microwave
background temperature to an accuracy of better than $10^{-4}$ is a
powerful confirmation of this principle.  As a point of further
support, the most recent large-scale redshift surveys$^{1)}$ indicate
a convergence to homogeneity also in the distribution of light.

However, on smaller scales inhomogeneities exist:  galaxies, cluster
of galaxies, voids and superclusters.  The isotropy of the microwave
background on smaller scales is an imprint of the homogeneity of the
matter distribution at the time of recombination.  Hence, it is
rather natural to work under the hypothesis that the present structure
of the Universe originates from the growth of initially small
cosmological perturbations.

At first glance, the theory of linear cosmological perturbations
appears straightforward.  Given a Friedmann-Robertson-Walker (FRW)
background model $\left (g_{\mu \nu}^{(0)}\, ,\, T_{\mu \nu}^{(0)}\right
)$ and small perturbations $\left (\delta g_{\mu \nu}\>,\> \delta T_{\mu
\nu}\right )$ of metric and energy-momentum tensor, we linearize the
Einstein equations
$$
G_{\mu \nu} = 8 \pi G T_{\mu \nu}\eqno(1)
$$
about the background solution to obtain
$$
\delta G_{\mu \nu} = 8 \pi G \delta T_{\mu \nu}\, .\eqno(2)
$$
The goal of the analysis of these equations is to find the time
dependence of the fractional density contrast $\delta \varepsilon
/\varepsilon$.

Linear cosmological perturbation theory was first developed by
Lifshitz in 1946, but prior to 1980 there was missing motivation for
any in depth study, the reason being that there was no causal theory
for the origin of fluctuations and hence no reason to study
perturbations except on length scales smaller than the Hubble radius
where Newtonian theory is adequate.

With the advent of inflationary Universe models, the situation changed
drastically.  As shown in Fig. 1, provided that the period of
inflation is sufficiently long, all scales of cosmological interest
originate inside the Hubble radius during the de Sitter phase.  Since
there is Hawking radiation in the de Sitter space$^{2)}$ with
temperature $T_H \sim H$, where $H$ is the expansion rate,
fluctuations are produced.  These
inhomogeneities evolve on scales much larger than the Hubble radius.
Hence, a general relativistic analysis is required.
\midinsert \vskip 8cm
\hsize=6in \raggedright
\noindent{\bf Figure 1:} Evolution of scales in the inflationary
Universe. The comoving wavelength $\lambda$ of a perturbation is
constant in comoving coordinates $x_c$. The Hubble radius increases
after inflation ($t > t_R$), but decreases exponentially during
inflation ($t < t_R$).
\endinsert

The Hawking radiation argument for the origin of perturbations given
above is too naive$^{3)}$.  The correct analysis uses the familiar
quantum particle production effects for quantum matter fields in an
expanding background, as applied to the scalar fields which drive
inflation.  If quantum fluctuations provide the seed perturbations for
structure in the Universe, then a quantum theory of cosmological
perturbations is required.

Hence, the key issues within the theory of cosmological perturbations
are
\item{ } -- to understand the growth of inhomogeneities on scales
larger than the Hubble radius,
\item{ } -- to develop a quantum theory of cosmological perturbations,
\item{ } -- to explain the quantum to classical transition for
fluctuations.

In this lecture, we will develop the classical and quantum theory of
cosmological perturbations, based on a recent comprehensive review
article$^{4)}$.  The issue of the quantum to classical transition will
not be addressed (see e.g. Ref. 5 for literature on this topic).
First, we will demonstrate that the analysis of the perturbation
equation (2) is not trivial:  there are gauge ambiguities, and the
best way to address this problem is to adopt an explicitly gauge
invariant formalism.
\medskip
\noindent \undertext{2.  Issues of Gauge}:
\par
In a general context, the gauge ambiguity can be described in two
ways.  In the passive view, we are given a space-time manifold $M$, a
physical quantity $Q$ defined on $M$, and some corresponding
coordinate function $^{(0)}Q(x)$ (in the above example,
$^{(0)}Q(\undertext{x}, t) = (t/t_0)^{1/2}$).  Let us now choose two
sets of coordinates $x$ and $\tilde x$ on $M$.  For the first choice,
the perturbation $\delta Q (p)$ of $Q$ at a point $p \in M$ is
defined as
$$
\delta Q (p) = Q (p) -\> {^{(0)}Q} (x (p))\, ,\eqno(3)
$$
whereas for the second choice
$$
\delta \tilde Q (p) = Q(p) - ^{(0)} Q (\tilde x (p))\, .\eqno(4)
$$
For small coordinate changes, the transformation $\delta Q (p)
\rightarrow \delta \tilde Q (p)$ is called a gauge transformation.

In the active view, we are given two manifolds, the space-time
manifold $M$ and an unperturbed background manifold $N$ with a fixed
coordinate choice.  To each physical quantity $Q$ on $M$ there is a
corresponding function $^{(0)}Q (x)$ on $N$.  Any coordinate choice
on $M$ corresponds to some map from $N$ to $M$ (see Fig. 2), and
hence to a different definition of $\delta Q(p)$ (see (3) and (4)).
\midinsert \vskip 8cm
\hsize=6in \raggedright
\noindent{\bf Figure 2:} The active view of a coordinate
transformation: the two mappings ${\cal D}$ and ${\tilde {\cal D}}$
from the background manifold $N$ to the physical manifold $M$ give
rise to two different coordinatizations of $M$ and hence to differing
definitions of perturbed quantities.
\endinsert

There are two approaches to the gauge problem.  One is to fix the
gauge, the other is to work in terms of gauge invariant variables.  We
will now argue that the use of gauge invariant variables has many
advantages.

Gravity is not the only theory with gauge ambiguities:
electromagnetism is another important example.  In electromagnetism we
can either work in terms of the gauge dependent potential $A_\mu$ or
in terms of the gauge invariant field strength tensor $F_{\mu \nu}$.
When using $A_\mu$, the homogeneous Maxwell equations are
automatically satisfied, and only the inhomogeneous ones need to be
solved explicitly.  Thus, working in terms of $A_\mu$ makes the
analysis easier in the sense that less equations must be solved.  The
disadvantage is that the variables have no direct physical meaning,
and that gauge artifacts like the Dirac string may appear.

In gravitational perturbation theory, however, no simplification of
the equations is\hfill\break
achieved by using gauge dependent variables.  Rather,
there are more equations and the analysis is more difficult.
The interpretational problems remain.  Hence, there is strong
motivation to adopt the gauge invariant formalism.

There is an additional reason for favoring the gauge invariant
approach over working in the usual gauge-synchronous gauge.  In
synchronous gauge there is a residual gauge  freedom which leads to
unphysical modes.  Although it is in principle possible to subtract
these modes, in practice there are formidable difficulties, especially
when working with approximate solutions.

Early attempts to develop a gauge invariant theory of cosmological
perturbations go back some time$^{6)}$.  The first completely gauge
invariant analysis was achieved by Gerlach and Sengupta$^{7)}$ and
Bardeen$^{8)}$.  This lagrangean approach was further developed  and
clarified in several papers$^{9)}$.  More recently, an alternative
Eulerian (or covariant) analysis has been developed in Refs. 10 and 11
and in many subsequent papers$^{12)}$.  The equations of Ref. 8 were
rederived in Ref. 13 using the Arnowitt-Deser-Misner approach.
For a recent review of the
classical and quantum theory of cosmological perturbations, the reader
is referred to Ref. 4.
\medskip
\noindent \undertext{3.  Classical Perturbations}:

\indent 3.1 {\it Formalism}
\par
There are three types of linear cosmological perturbations:  scalar,
vector and tensor modes.  The names refer to the way in which the
modes transform under background space coordinate transformations (see e.g.
Ref. 14). Tensor modes are
gravitational waves, vector perturbations correspond to rotation and
do not grow in time, and only the scalar modes couple (via the
Einstein equations) to energy density and pressure.  Hence, we shall
restrict our attention to scalar type cosmological perturbations.

The first step in the analysis of cosmological perturbations is to
identify the gauge invariant combinations of $\delta g_{\mu \nu}$.
The general scalar metric perturbation can be written
in terms of four scalar functions $\phi,\>\psi,\> B$ and $E$
$$
\delta g_{\mu \nu} = a^2 \left(\matrix{2\phi & - B_{, i}\hfill\cr
-B_{, i}\qquad &2(\psi \delta_{ij} - E_{, ij})\cr}\right )\,
.\eqno(5)
$$
For simplicity, we have restricted our attention to the case of a
spatially flat background.  The following gauge transformations
preserve the scalar character of $\delta g_{\mu \nu}$:
$$
\eqalign{
\tilde\eta &= \eta + \xi^0\cr
{\tilde x}^i &= x^i + \gamma^{ij} \xi_{, j}}\eqno(6)
$$
where $\xi^0$ and $\xi$ are functions of space and time.
It is not hard to check that the induced changes of $\phi, \psi, B$
and $E$ are
$$\eqalign{
{\tilde \phi} &= \phi - {a^\prime\over {a}}\> \xi^0 - {\xi^0}^\prime\cr
{\tilde \psi} &= \psi + {a^\prime\over {a}}\> \xi^0\cr
{\tilde B} &= B + \xi^0 - \xi^\prime\cr
{\tilde E} &= E-\xi\, ,}\eqno(7)
$$
where a prime denotes the derivative with respect to conformal time.
Now it is a simple exercise in linear algebra to find a basis of gauge
invariant variables. A convenient choice is
$$\eqalign{
\Phi &= \phi + a^{-1} [(B-E^\prime\,)a]^\prime\cr
\Psi &= \psi - {a^\prime\over {a}}\> (B-E^\prime\,)\, .}\eqno(8)
$$
Note that in longitudinal gauge $(B=E=0)$ the gauge invariant
variables become $\Phi = \phi$ and $\Psi = \psi$.
\par
The second step of our analysis is to derive the equations of motion
for the gauge invariant variables.  In principle, this is
straightforward.  The linearized Einstein equations (2) are
conveniently combined to yield equations for $\Phi$ and $\Psi$.  In
practice, this computation is rather tedious unless a clever
procedure is chosen.  It is simplest$^{4)}$ to consider first the
transformation of the perturbation $\delta G_{\mu \nu}$ of the Einstein
tensor under (6), and to determine gauge invariant combinations
(labelled with superscript $(gi)$):
$$\eqalign{
\delta {G_0^0}^{(gi)} &= \delta G_0^0 +\> {^{(0)}G}_0^{0 \prime}\, (B-
E^\prime\, )\cr
\delta {G_i^0}^{(gi)} &= \delta G_i^0 + \left ( ^{(0)}G_0^0 - {1\over
3}\> \>^{(0)}G_k^k\right ) (B-E^\prime\, )_{, i}\cr
\delta {G_j^i}^{(gi)} &= \delta G_j^i +\> ^{(0)}G_j^{i\prime}\>(B-
E^\prime\, )}\eqno(9)
$$
where the background Einstein tensor elements are $^{(0)}G_\nu^\mu$.
Evidently, the analogous combinations of $\delta T_\nu^\mu$ are gauge
invariant.  Thus, the linearized Einstein equations can be written as
$$
\delta {G_\nu^\mu}^{(gi)} = 8 \pi G\>{\delta T_\nu^\mu}^{(gi)}\,
.\eqno(10)
$$
In this form, all the gauge dependence automatically drops out, and we
obtain the following set of equations written exclusively in terms of
gauge invariant variables:
$$\eqalign{
-3{\cal H} ({\cal H} \Phi + \Psi^\prime\, ) + \nabla^2 \Psi &= 4\pi G
a^2 {\delta T_0^0}^{(gi)}\cr
\>\>\>\>\>({\cal H} \Phi + \Psi^\prime\, )_{, i} &= 4\pi G a^2 {\delta
T_i^0}^{(gi)}\cr
\Bigl [ (2{\cal H}^\prime + {\cal H}^2) \Phi + {\cal H} \Phi^\prime +
\Psi^{\prime \prime}\Bigr. &+\Bigl.
2{\cal H} \Psi^\prime + {1\over 2}\> \nabla^2
 D\Bigr ]
\delta_j^i - {1\over 2} \gamma^{ik} D_{, kj}\cr
&= - 4\pi G a^2 {\delta T_j^i}^{(gi)}\, ,}\eqno(11)
$$
where $D = \Phi - \Psi$ and ${\cal H} = a^\prime\, /a$.

An alternative way to derive the above equations$^{4)}$ is to work in
longitudinal gauge and at the end replace $\phi$ and $\psi$ by $\Phi$
and $\Psi$ respectively (and similarly for the matter variables).

\indent 3.2 {\it Applications}
\par
As a first application of the classical theory of cosmological
perturbations we shall consider the example of perfect fluid matter
given by the energy-momentum tensor
$$
T_\beta^\alpha = (\varepsilon + p) u^\alpha u_\beta -
p\delta_\beta^\alpha\, ,\eqno(12)
$$
$\varepsilon$ and $p$ being energy density and pressure respectively,
and $u^\alpha$ the four velocity vector of the fluid.  In general, the
pressure is a function of both $\varepsilon$ and entropy per baryon
$s$, and hence
$$
\delta p = c_s^2 \delta \varepsilon + \tau \delta s\, ,\eqno(13)
$$
with $c_s$ being the speed of sound.  If $\tau = 0$, we have a pure
adiabatic perturbation.

The perturbation of $T_\beta^\alpha$ is given by
$$\eqalign{
\delta T_0^0 &= \delta \varepsilon\cr
\delta T_i^0 &= (\varepsilon_0 + p_0) a^{-1} \delta u_i\cr
\delta T_j^i &= - \delta p \delta_j^i\, ,}\eqno(14)
$$
where subscripts denote background quantities.  Since $\delta T_j^i$
is diagonal, it follows immediately from the third equation in (11)
that $\Phi = \Psi$ . This, in turn, leads to a significant simplification of
the equations
of motion for the gauge invariant variables.  From the
first equation in (11), we obtain
$$
\nabla^2 \Phi - 3 {\cal H} \Phi^\prime - 3 {\cal H}^2 \Phi = 4 \pi
Ga^2 \delta \varepsilon^{(gi)}\, .\eqno(15)
$$
This is a generalization of the Poisson equation to which it reduces
in the Newtonian limit; and hence we call $\Phi$ the relativistic
potential.

Equations (11) can be combined to yield the following
second order equation of motion for $\Phi$:
$$
\Phi^{\prime\prime} + 3{\cal H} \left (1 + c_s^2\right ) \Phi^\prime -
c_s^2 \nabla^2 \Phi + \left [ 2{\cal H}^\prime + (1 + 3c_s^2 ) {\cal
H}^2 \right ]\Phi = 4\pi Ga^2 \tau \delta s\eqno(16)
$$
For adiabatic perturbations, the source term vanishes.  On scales
larger than the Hubble radius, the spatial gradients can be neglected.
Under these conditions, equation (16) can be recast as a
``conservation law"
$$
{\dot \zeta} = 0\eqno(17)
$$
where the dot denotes the derivative with repect to physical time and
$$
\zeta = {2\over 3}\> {H^{-1} {\dot \Phi} + \Phi\over {1+w}} +
\Phi\eqno(18)
$$
with $w = p/\varepsilon$.  The quantity $\zeta$ was first introduced in
Ref. 15 (see also Ref. 16).  The above conservation law is easily
applicable to many interesting issues.  First, we note that if the
equation of state is constant, then $\Phi$ remains constant (the
second solution of (17) is a decaying mode).  However, during a phase
transition $w$ may change by a large factor.  In this case, equation
(17) implies that the relativistic potential $\Phi$ will also change
by a large factor.  This is one of the key points in the computation
of density perturbations from inflation$^{17)}$.

To correctly describe fluctuations from inflation, we must consider a
second application of the classical theory of cosmological
perturbations, namely a model with scalar field matter.  The matter
action is
$$
S_m = \int d^4 x \sqrt{-g} \left\{ {1\over 2}\> {\varphi^;}^\alpha
\varphi_{;^\alpha} - V (\varphi)\right\}\, ,\eqno(19)
$$
semicolons denoting the covariant derivative.  The induced energy-
momentum tensor of the scalar field $\varphi$ is
$$
T_\beta^\alpha = {\varphi^;}^\alpha \varphi_{; \beta} - \left\{
{1\over 2}\> \varphi^{;\gamma} \varphi_{; \gamma} - V (\varphi)\right\}
\delta^\alpha_\beta\, .\eqno(20)
$$
If we expand $\varphi (\undertext{x},t)$ about a homogeneous
background field $\varphi_0 (t)$
$$
\varphi ({\undertext x}, t) = \varphi_0 (t) + \delta \varphi
({\undertext x}, t)\, ,\eqno(21)
$$
then the perturbation of $T_\beta^\alpha$ at the linearized level becomes
$$\eqalign{
\delta T_0^0 &= a^{-2} \left\{ -\varphi_0^{\prime 2} \phi +
\varphi_0^\prime \delta \varphi^\prime + V_{, \varphi} a^2 \delta
\varphi \right\}\cr
\delta T_i^0 &= a^{-2} \varphi_0^\prime \delta \varphi_{, i}\cr
\delta T_j^i &= a^{-2} \left\{ \varphi_0^{\prime 2} \phi -
\varphi_0^\prime \delta \varphi^\prime + V_{, \varphi} a^2 \delta
\varphi \right\} \delta_j^i\, .}\eqno(22)
$$
As in the case of a perfect fluid, $\delta T_j^i$ is diagonal and
hence $\Phi = \Psi$

Inserting thios result and (22) into the general equations (11) and combining
the resulting differential equations, we obtain the
following second order equation for $\Phi$
$$
\Phi^{\prime \prime} + 2 \left ({\cal H} -
{\varphi_0^{\prime\prime}\over {\varphi_0^\prime}}\right ) \Phi^\prime -
\nabla^2 \Phi + 2 \left ({\cal H}^\prime -
{\varphi_0^{\prime\prime}\over {\varphi_0^\prime}}\, {\cal H}\right
)\Phi = 0 \, .\eqno(23)
$$
Since for a scalar field
$$
1 + w = {{\dot \varphi}_0^2\over {\varepsilon}}\, ,\eqno(24)
$$
we can, like for perfect fluid matter, rewrite (23) as a
``conservation law" identical to (17) and (18) when considering scales
much larger than the Hubble radius.

\indent 3.3 {\it Fluctuations in Inflationary Cosmology}
\par
To demonstrate how easy it is to apply the gauge invariant theory of
cosmology perturbations, we shall consider the evolution of
fluctuations in inflationary Universe models$^{15-19)}$. We first note
from (11) that on scales smaller than the Hubble radius
$$
\Phi = - {3\over 2} \left ({aH\over {k}}\right )^2\> \left ({\delta
\varepsilon\over {\varepsilon}}\right )^{(gi)}\, .\eqno(25)
$$
The calculation of density perturbations proceeds as follows:  by
evaluating (25) at the time $t_i (k)$ (see Fig. 1) when the wavelength
under consideration leaves the Hubble radius, we determine the initial
value of $\Phi, \Phi (t_i (k))$.  By integrating (17) and using the
fact that ${\dot \Phi}$ vanishes at both $t_i (k)$ and $t_f (k)$, the
value of $\Phi$ at the time $t_f (k)$ when the scale reenters the
Hubble radius can be calculated with the result
$$
\Phi (t_f (k)) = {1+ w(t_f)\over {{5\over 3} + w (t_f)}}\>\> {2\over
3}\>\> {\Phi (t_i)\over {1 + w (t_i)}} \equiv \alpha\> {\Phi (t_i)\over
{1 + w(t_i)}}\, .\eqno(26)
$$
The coefficient $\alpha$ is $4/9$ for $t_f$ in the radiation dominated
phase and $\alpha = 2/5$ during matter domination.  Using (25), the
value of $\Phi (t_f)$ determines the late time value of the amplitude
of the fractional density perturbation.

In order to evaluate the amplitude of perturbations (26) in inflationary
Universe models, a quantum analysis of the generation of fluctuations is
required. However, already a rough order of magnitude estimate yields
interesting results. From (25) and taking the energy density perturbation to be
given by the Hawking temperature (i.e. $\delta \varepsilon \sim H^4$) we obtain
$$
\Phi(t_i) \sim {H^4 \over \varepsilon} \ll 1.\eqno(27)
$$
Inflation is driven by a scalar field $\varphi$. During inflation, the equation
of state is dominated by the potential energy density of $\varphi$. However,
$\varphi$ is rolling and therefore
$$
1 + w(t_i) = {{{\dot \varphi}^2} \over \varepsilon}.\eqno(28)
$$
On dimensional grounds, ${\dot \varphi}^2 \sim H^4$ and hence $\Phi(t_f) \sim
1$. We conclude that the change in the equation of state leads to a drastic
amplification of the initial quantum fluctuations. Successful models of galaxy
formation require $\Phi(t_f) \sim 10^{-4}$. Thus, without careful adjustment of
parameters, inflationary Universe models predict too large
perturbations$^{18)}$.
\endpage
\noindent \undertext{4.  Quantum Perturbations}:

\indent 4.1 {\it Motivation}
\par
The classical analysis of fluctuations in inflationary Universe models
gives good insight into why initially tiny inhomogeneities are
amplified by a large factor between when they are produced in the de
Sitter phase and when they reenter the Hubble radius at late times.
It is, however, only a quantum analysis which explains the origin of
these perturbations.  It is vacuum quantum fluctuations which are the
source of the classical inhomogeneities which form the seeds for
galaxy and cluster formation.

A second motivation for considering the quantum theory of cosmological
perturbations comes from the general problem of particle production in
expanding background space-times.  The usual$^{20)}$ treatment which
is based on quantizing matter fields on an unperturbed cosmological
background is inconsistent since matter fluctuations are intrinsically
coupled to metric perturbations via the Einstein equations.  Hence, we
need to quantize metric and matter fluctuations in a unified way.

In fact, the quantization of linear cosmological fluctuations is not
more complicated than the well known quantization of matter fields in
an external background:  it is a straightforward application of
canonical quantization$^{20)}$. In synchronous gauge, the quantization of
fluctuations was first discussed in Ref. 17.

Since we only wish to quantize the physical degrees of freedom, it is
advantageous to use the gauge-invariant formalism.  Since this method
reduces the number of degrees of freedom, it also leads to a
substantial simplification of the analysis.

The first step in deriving the quantum theory of cosmological
perturbations$^{21, 22)}$ is to determine the action for the
fluctuations in terms of the gauge invariant variables.  In general,
it would be wrong to simply start from the classical equation of
motion for perturbations and interpret it directly as an operator
equation.  This would lead to wrong canonical momenta and to a wrong
normalization of the field operator$^{4, 23)}$.

\indent 4.2 {\it Formalism}
\par
In the following, we shall briefly summarize the quantum theory of
cosmological perturbations.  For simplicity, only models with scalar
field matter will be considered.  For hydrodynamical matter the
analysis is similar$^{4)}$.  The formalism also applies to highter
derivative gravity theories$^{4, 24)}$.  We will follow the method of
Ref. 21 (see Ref. 4 for more details).

The first and most involved step in quantizing cosmological
perturbations is to write the action for fluctuations in terms of
gauge invariant variables only.  We start from the action
$$
S = \int d^4 x \sqrt{-g}\> R + S_m\, ,\eqno(29)
$$
where $S_m$ is the action for the scalar field $\varphi$.  Next, we
insert into (29) the expansion of $g_{\mu \nu}$ and $\varphi$ about a
homogeneous background solution $g_{\mu \nu}^{(0)}$ and $\varphi_0$
$$\eqalign{
g_{\mu \nu} &= g_{\mu \nu}^{(0)} + \delta g_{\mu \nu}\cr
\varphi &= \varphi_0 + \delta \varphi}\eqno(30)
$$
and expand the result in terms of powers of small quantities to find
$$
S = S_0 + \delta_2 S\eqno(31)
$$
where $S_0$ is the action of the background solution and $\delta_2 S$
is quadratic in perturbation variables (the linear terms vanish
because we are expanding about a solution of the equations of motion).
We now use the constraint equations to simplify the action and drop
total derivative terms.  After a significant amount of algebra one
obtains the following very simple form of $\delta_2 S$:
$$
\delta_2 S = {1\over 2} \int d^4 x \left\{v^{\prime^2} - v_{, i}\> v_{,
j}\> \delta^{ij} + {z^{\prime\prime}\over {z}} v^2\right\}\eqno(32)
$$
where $v$ is a gauge invariant combination of matter and metric
perturbations
$$
v = a \left (\delta \varphi^{(gi)} + {\varphi_0^\prime\over {\cal H}}
\Phi \right )\eqno(33)
$$
and
$$
z = {a \varphi_0^\prime\over {\cal H}}.\eqno(34)
$$

The result (32) has the same form as the action of a simple scalar
field with time dependent square mass $- z^{\prime\prime} / z$.  Note
that although the details of the reduction of the action are somewhat
involved, the final result is no surprise.  We have seen in section 3
that for scalar field matter there is only one independent gauge
invariant metric perturbation variable.  Via the Einstein equations
this variable is coupled to the gauge invariant matter fluctuations.
Thus, this is only one independent variable  which expresses in a
unified manner both matter and metric perturbations.

{}From this point on, the quantization prescription is straightforward
canonical quantization.  From $\delta_2 S$ we can immediately write
down the canonical momenta.  After imposing the canonical commutation
relations, we expand the operator $\hat v$ corresponding  to the
classical field $v$ in terms of creation and annihilation operators
$a_k^+$ and $a_k^{-}$:
$$
{\hat v} = {1\over 2} {1\over {(2 \pi)^{3/2}}} \int d^3 k \left [
e^{ikx} v_k^\ast (\eta) a_k^{-} + e^{- ikx} v_k (\eta) a_k^+\right ]\,
.\eqno(35)
$$
The mode functions $v_k (\eta)$ satisfy the equation
$$
v_k^{\prime \prime} + \left ( k^2 - {z^{\prime \prime}\over {z}}\right
) v_k = 0\, .\eqno(36)
$$

Since (36) is a harmonic oscillator equation with time dependent mass,
there will be quantum particle production$^{20)}$.  Modes of (36)
which have positive frequency at some initial time $t_0$ are no longer
pure positive frequency at a later time $t_1 > t_0$.  This leads to
time dependence of expectation values of physical operators.  For
example, if $\vert \psi_0 >$ is the vacuum state at time $t_0$, and
$N_k (t_1) = a_k^+ (t_1) a_k^{-} (t_1)$ is the number operator at time
$t_1$ defined in terms of the operator coefficients of the positive
frequency modes at time $t_1$, then
$$
< \psi_0 | N_k (t_1) |\psi_0 > \neq 0\, .\eqno(37)
$$

The final step is to compute the expectation values of the operators
which determine the r.m.s. mass fluctuation.  If $\delta M / M (k)$ is the
r.m.s. mass perturbation inside a sphere of radius $k^{-1}$, then
$$
\left ({\delta M\over M}\right )^2 (k) \sim k^3 \left ({\delta
\varepsilon\over \varepsilon}\right )^2 (k)^{(gi)} = k^3 {<\psi_0 |
|\delta\varepsilon^{(gi)} (k)|^2 |\psi_0 >\over {\varepsilon^2}}\,
,\eqno(38)
$$
where in the final step we have replaced the classical perturbation by
the expectation value of the quantum operator evaluated in the vacuum
state $| \psi_0 >$ at the beginning of inflation.  This prescription
for taking the quantum to classical transition has been discussed in
Ref. 5 and references therein.  When evaluated at the time of Hubble
radius crossing $t_f (k)$ (see Fig. 1), then using the relationship
(25) between $\delta \varepsilon^{(gi)}$ and $\Phi$ one obtains
$$
\left ( {\delta M\over M}\right )^2 (k , t_f (k)) \sim k^3 <\psi_0 |
|\Phi (k) |^2 |\psi_0 >\, .\eqno(39)
$$
In turn, the gauge invariant potential $\Phi$ is related to the
variable $v$ by
$$
k^2 \Phi = - 4 \pi G\>\> {\varphi_0^{\prime^2}\over {\cal H}} \left (
{v\over z}\right )^\prime\, .\eqno(40)
$$
Hence, the computation of the expectation value in (39) reduces to a
straightforward evaluation of the expectation value of $v^2$.

Combining (39) and (40), we find
$$
\left ( {\delta M\over M}\right )^2 (k , t_f (k)) \sim {1\over {4
\pi^2}}\>\> {\varphi_0^{\prime^2}\over {a^2}}\>\> k^3 | u_k (t_f
(k))|^2\eqno(41)
$$
where $u_k (\eta)$ are proportional to the expansion coefficients of
the operator $\hat \Phi$:
$$
{\hat \Phi} ({\undertext x}, \eta) = {1\over {\sqrt{2}}} {1\over
{(2\pi)^{3/2}}} {\varphi_0^\prime\over  a} \int d^3 k \left [ u_k^\ast
(\eta) e^{i{\undertext k}\cdot {\undertext x}} a_k^{-} + u_k (\eta) e^{-i
 {\undertext k}\cdot {\undertext x}} a_k^+\right ]\, .\eqno(42)
$$
Equation (41) relates the r.m.s. mass perturbation resulting from
quantum vacuum fluctuations to the solution $u_k(\eta)$ of the classical
equation
of motion.  At this point, we have established a consistent unified
treatment of generation and evolution of cosmological perturbations.

Evaluating (41) for a model of chaotic inflation$^{25)}$ with potential
$$
V(\varphi) = {\lambda \over n} \varphi^n \eqno(43)
$$
yields
$$
{{\delta M} \over M} (k, t_f(k)) \sim {\lambda^{1/2} \over {m_{pl}}} \bigl(
{{2n} \over {3 l^2}} \bigr)^{n/4 - 1/2} [ln({{k_{\gamma}} \over k})]^{n/4 +
1/2},\eqno(44)
$$
where $l$ is the Planck length and $k_{\gamma}$ is the characteristic
wavenumber of the cosmic microwave background. For $n = 2$ (and setting
$\lambda = m^2$), a correct value of ${{\delta M} \over M}$ for galaxy
formation requires $m \sim 10^{-6} m_{pl}$, for $n = 4$ the requirement is
$\lambda \sim 10^{-12}$. As mentioned at the end of Section 3, inflation thus
requires careful adjustment of parameters if it is to give the required value
of density perturbations.

This completes our brief survey of the quantum theory of cosmological
perturbations.  Although the formalism has been developed for scalar
field matter (and in particular applied to quantum fluctuations in
inflationary Universe models), it is much more general.  Also for
hydrodynamical matter and in higher derivative gravity theories, the
action for perturbations can be reduced to a form like (32), and the
quantization then proceeds as in the example discussed$^{4)}$.
\medskip
\noindent \undertext{5.  Conclusions and Discussion}:
\par
We have summarized the gauge invariant theory of classical and
quantum cosmological perturbations.  It allows a consistent unified
treatment of the generation and evolution of linearized fluctuations
in  inflationary Universe models.

In Section 2 we argued that a gauge invariant analysis of classical
perturbations is physically unambiguous and technically
straightforward.  It eliminates the gauge ambiguities associated with
gauge-dependent approaches.  The coordinate approach presented here is
probably the most simple way of deriving the equations of motion for
the gauge invariant gravitational potential.  It is action based and
hence allows standard canonical quantization.  A gauge invariant
analysis of the quantum theory implies that only physical degrees of
freedom are quantized.

The formalism presented here is practical and can easily be applied to
problems of real cosmological interest.  Already the classical theory
leads to a useful ``conservation law" (see (17) and (18)) which allows
us to track the amplitude of perturbations on scales much larger than
the Hubble radius in a very simple manner.

The quantum theory of cosmological perturbations relates the
expectation values of two-point functions which determine the r.m.s.
mass fluctuations to mode functions which obey the classical equations
of motion for the gauge invariant gravitational potential.  This
allows a unified analysis of the generation and evolution of density
perturbations in inflationary Universe models (see (41)). A general formula for
the amplitude of the resulting fluctuations is given.

In Ref. 4 we have performed detailed calculations of the spectrum of
density perturbations in models with scalar field matter,
hydrodynamical matter, and in higher derivative theories of gravity.
The formalism also allows a discussion of entropy perturbations, it
can be  used to yield a simple proportionality between the microwave
background temperature anisotropies and the gravitational potential
$\Phi$, and it can be  applied to the generation and evolution of
gravitational waves.
\medskip
\noindent \undertext{Acknowledgements}:
\par
One of us (R.B.) is grateful to Professors K.-I. Maeda and K. Sato for the
invitation to speak at this Yamada conference and for their hospitality in
Tokyo. R.B.
is supported in part by DOE grant DE- FG02-91-ER40688, Task A.
\medskip
\REF\one{S. Shechtman et al., `Strip-Mining the Southern Sky: Skratching the
Surface', CFA preprint No. 3385 (1992).}
\REF\two{E. Lifshitz, {\it Zh. Eksp. Teor. Fiz.} {\bf 16}, 587
(1946).}
\REF\three{See e.g. discussion session  on inflation at ICGC-87, in
``Highlights in Gravitation and Cosmology", ed. by B. Iyer, et. al.
(Cambridge Univ. Press, Cambridge, 1988).}
\REF\four{V. Mukhanov, H. Feldman and R. Brandenberger, {\it Phys.
Rep.} {\bf 215}, 203 (1992).}
\REF\five{M. Sakagami, {\it Prog. Theor. Phys.} {\bf 79}, 443 (1988);\nextline
R. Brandenberger, R. Laflamme and M. Mijic, {\it Phys. Scripta} {\bf
T36}, 265 (1991).}
\REF\six{S. Hawking, {\it Ap. J.} {\bf 145}, 544 (1966);\nextline D. Olson,
{\it Phys. Rev.} {\bf D14}, 327 (1976).}
\REF\seven{U. Gerlach and U. Sengupta, {\it Phys. Rev.} {\bf D18}, 1789
(1978).}
\REF\eight{J. Bardeen, {\it Phys. Rev.} {\bf D22}, 1882 (1980).}
\REF\nine{R. Brandenberger, R. Kahn and W. Press, {\it Phys. Rev.}
{\bf D28}, 1809 (1983);\nextline H. Kodama and M. Sasaki, {\it Prog. Theor.
Phys. Suppl.} {\bf 78}, 1 (1984);\nextline G. Chibisov and V. Mukhanov,
``Theory of Relativistic Potential:  Cosmological Perturbations",
Preprint No. 154 of P.N. Lebedev Physical Institute (1983).}
\REF\ten{D. Lyth and M. Mukherjee, {\it Phys. Rev.} {\bf D38}, 485
(1988).}
\REF\eleven{G. Ellis and M. Bruni, {\it Phys. Rev.} {\bf D40}, 1804
(1989).}
\REF\twelve{G. Ellis, J. Hwang and M. Bruni, {\it Phys. Rev.} {\bf
D40}, 1919 (1989);\nextline J. Hwang and E. Vishniac, {\it Ap. J.} {\bf 353}, 1
(1990);\nextline G. Ellis, M. Bruni and J. Hwang, {\it Phys. Rev.} {\bf D42},
1035 (1990).}
\REF\thirteen{R. Durrer and N. Straumann, {\it Helv. Phys. Acta} {\bf
61}, 1027 (1988).}
\REF\fourteen{J. Stewart, {\it Class. Quant. Grav.} {\bf 7}, 1169
(1990).}
\REF\sfifteen{J. Bardeen, P. Steinhardt and M. Turner, {\it Phys. Rev.}
{\bf D28}, 679 (1983).}
\REF\sixteen{R. Brandenberger and R. Kahn, {\it Phys. Rev.} {\bf
D29}, 2175 (1984);\nextline D. Lyth, {\it Phys. Rev.} {\bf D31}, 1792 (1985).}
\REF\seventeen{G. Chibisov and V. Mukhanov, ``Galaxy Formation and
Phonons", Lebedev Physical Institute Preprint No. 162 (1980);\nextline G.
Chibisov and V. Mukhanov, {\it Mon. Not. R. astron. Soc.} {\bf 200},
535 (1982);\nextline V. Lukash, {\it Pisma Zh. Eksp. Teor. Fiz.} {\bf 31}, 631
(1980);\nextline V. Lukash, {\it Zh. Eksp. Teor. Fiz.} {\bf 79}, 1601 (1980).}
\REF\eighteen{S. Hawking, {\it Phys. Lett.} {\bf 115B}, 295 (1982);\nextline A.
Starobinsky, {\it Phys. Lett.} {\bf 117B}, 175 (1982);\nextline A. Guth and S.-
Y. Pi, {\it Phys. Rev. Lett.} {\bf 49}, 1110 (1982).}
\REF\nineteen{V. Mukhanov, {\it Pisma Zh. Eksp. Teor. Fiz.} {\bf 41},
402 (1985).}
\REF\twenty{N. Birrell and P. Davies, ``Quantum Fields in Curved
Space", (Cambridge University Press, Cambridge, 1982).}
\REF\twentyone{V. Mukhanov, {\it Zh. Eksp. Teor. Fiz.} {\bf 94}, 1
(1988).}
\REF\twentytwo{M. Sasaki, {\it Prog. Theor. Phys.} {\bf 76}, 1036
(1986).}
\REF\twentythree{N. Deruelle, C. Gundlach and D. Polarski, {\it Class.
Quant. Grav.} {\bf 9}, 137 (1992).}
\REF\twentyfour{V. Mukhanov, {\it Phys. Lett.} {\bf 218B}, 17
(1989).}
\REF\twentytwo{A. Linde, {\it Phys. Lett.} {\bf 129B}, 177 (1983).}
\refout
\end